\documentclass[prl,twocolumn,amsmath,showpacs]{revtex4}
\input {epsf.sty}
\usepackage{epsfig}
\usepackage{graphicx}
\usepackage{amsmath,amsthm,amssymb,latexsym,amsfonts,times}
\usepackage{bm}

\begin{document}
\title{Magneto-Acoustic Spectroscopy in Superfluid $^3$He-B}
\author{J.P. Davis, H. Choi, J. Pollanen, and W.P. Halperin}
\affiliation{Department of Physics and Astronomy, Northwestern 
University, Evanston,
Illinois 60208}

\date{Version \today}


\begin{abstract}  We have used the recently discovered acoustic 
Faraday effect in
superfluid $^3$He to perform high resolution spectroscopy of an 
excited state of the
superfluid  condensate.   With acoustic cavity interferometry we 
measure the rotation of
the plane of polarization of a  transverse sound wave propagating in 
the direction of
magnetic field from which we determine the Zeeman energy of the 
excited state. We
interpret the Land$\acute{e}$ $g$-factor, combined with the 
zero-field energies of the
state, using the theory of Sauls and Serene to calculate the strength 
of $f$-wave
interactions in $^3$He.
\end{abstract}

\maketitle

\vspace{11pt} Magneto-acoustic effects are less commonly known than 
their magneto-optical
counterparts,  but similar phenomena occur with transverse sound if 
there is significant
magneto-elastic coupling.  An acoustic Faraday effect (AFE) was first 
predicted for
ferromagnetic crystals by Kittel
\cite{Kit58} and subsequently  was observed in magnetically ordered materials
\cite{Mat62,Boi71}.  It may also occur  in the vortex lattice of 
high-$T_c$ superconductors
\cite{Dom95}.  At low temperature in the superfluid state of $^3$He, 
the spontaneously
broken relative spin-orbit symmetry of the B-phase provides the 
mechanism for such a
coupling.  An AFE was predicted by Moores and Sauls \cite{Moo93} and 
observed by Lee {\it
et al.} \cite{Lee99}  proving that transverse sound waves exist in 
superfluid $^3$He. It is
of special note that  this is the only known case where transverse 
waves propagate in a
fluid.  With recent improvements in acoustic cavity  techniques 
\cite{Dav06} we have
achieved sufficiently high spectroscopic resolution over a wide range 
of  frequency to
measure the AFE and study the excited states of the condensate as a 
function of pressure.
The pressure is an essential variable that permits one to vary the  strength of
quasiparticle and pairing interactions.  Using the theory of Sauls and  Serene
\cite{Sau82} we relate our Faraday rotation results to these interactions.

The acoustic Faraday effect requires two coupled modes; a 
transverse-sound mode,
linearly-dispersive in the absence of magneto-elastic coupling, with 
phase velocity
$c = \omega/q$ for frequency $\omega$ at wave vector $q$.  The second 
mode must be
magnetically active, $\omega^{2} =
\Omega_{0}^{2}(H) + b q^{2}$, having a magnetic field dependence in 
the long wavelength
limit and quadratic dispersion. For ordered materials
$\Omega_{0}$ is the ferro- or  antiferro-magnetic resonance 
frequency, proportional to the
internal local field and shifted linearly by an applied field,
$H$.  For superfluid $^3$He the magnetic mode, $\Omega_{2^-}$, is an 
excited state of  the
superfluid condensate, having total angular momentum
$J=2$, called the {\it imaginary} squashing collective mode (ISQ), a 
reference to the
nature  of the order parameter distortions that characterize it, and 
to distinguish it
from another $J=2$ mode that involves only {\it real} components of 
the order parameter
\cite{Hal90}. The linear  Zeeman splitting of the ISQ-mode is the 
source of the magnetic
field dependence, which leads to  different couplings for right and 
left circularly
polarized transverse sound, i.e. acoustic circular birefringence.

\begin{figure}[b]
     \centering
\centerline{\includegraphics[width=2.8in]{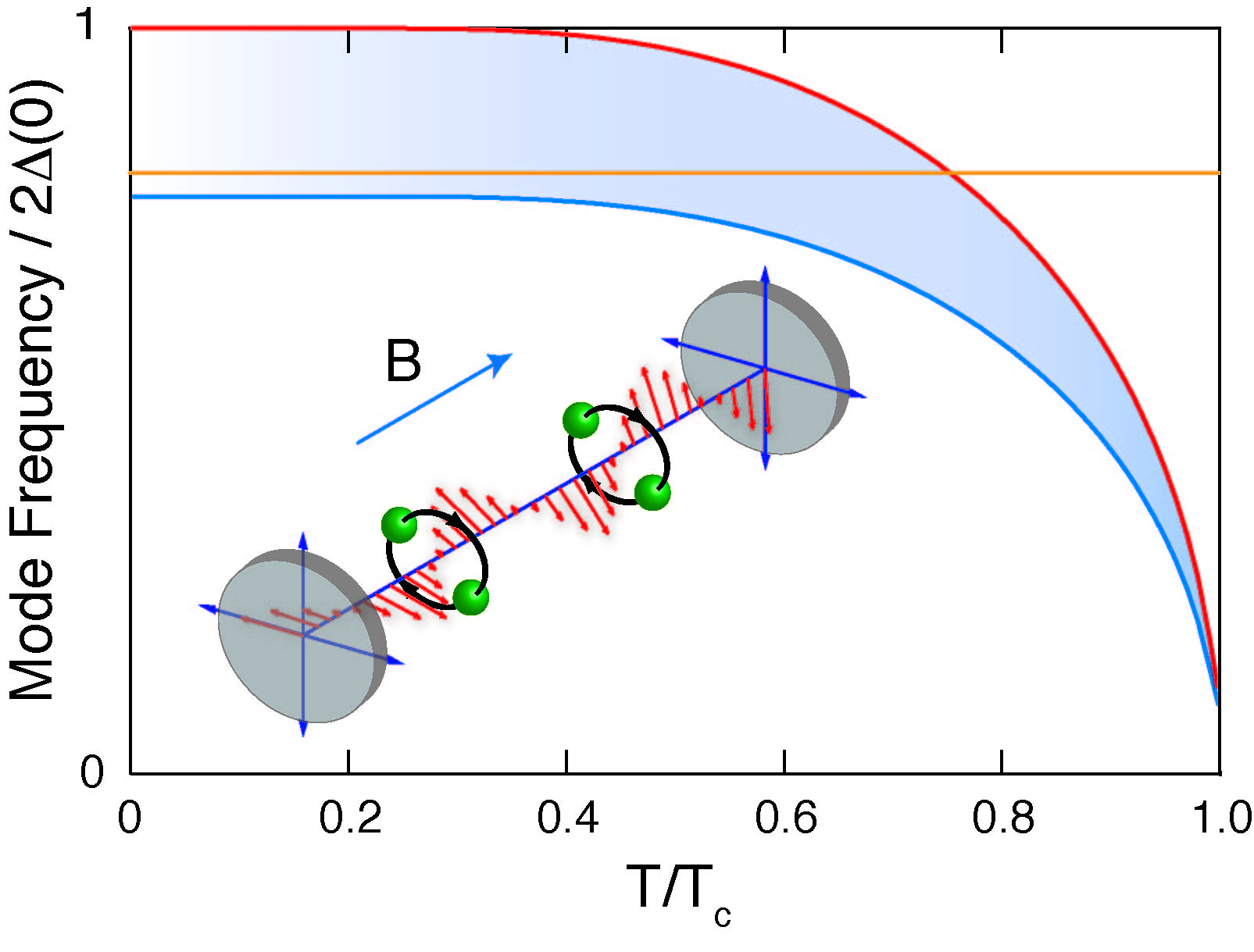}}
\caption {\label{fig1}(color online).  The ISQ-mode frequency (lower 
blue curve) and
pair-breaking  (upper red curve) relative to the zero temperature gap 
as a function of
temperature.  Propagation of transverse sound, generated at constant 
frequency (orange
line), is observed in the shaded (blue) region.  The cartoon shows 
the acoustic  Faraday
effect, in which linearly polarized transverse waves (red arrows) 
are generated by a
piezoelectric  transducer and rotated by a magnetic field.  The green spheres
represent $^3$He Cooper  pairs coupling to transverse sound.}
\end{figure}

The coupled-mode dispersion relation for superfluid $^3$He in the 
long wavelength limit
\cite{Moo93,Sau99} can be written as,
\begin{equation}\label{dispersion}
     \frac{\omega^2}{q^2 v_{F}^2} = \Lambda_{0} +
      \Lambda_{2^{-}}\frac{\omega^{2}}{\omega^{2} -
\Omega_{2^{-}}^{2}-\frac{2}{5}q^{2}v_{F}^{2}- 2m_{J}g_{2^{-}}
\gamma_{eff} H\omega},
\end{equation}
  where $\gamma_{eff}$ is the effective gyromagnetic ratio of $^3$He, 
$m_{J}$ is the total
angular momentum substate quantum number, and $v_{F}$ is the Fermi 
velocity. The ISQ-mode
frequency follows the temperature and pressure dependence of the energy gap,
$\Delta(T,P)$, shown in Fig.~\ref{fig1}, or more precisely 
$\Omega_{2^{-}}^{2} =
a_{2^{-}}^2\Delta^2$, where $a_{2^{-}} \approx \sqrt{12/5}$.  The first term in
Eq.~\ref{dispersion} is a quasiparticle background, $\Lambda_{0} =
\frac{F_1^{s}}{15}(1-\lambda)(1+\frac{F_{2}^{s}}{5})/(1+\lambda\frac{ 
F_{2}^{s}}{5})$.
The magneto-acoustic coupling strength is
$\Lambda_{2^{-}} = \frac{2F_1^{s}}{75}\lambda
      (1+\frac{F_{2}^{s}}{5})^{2})/(1+\lambda\frac{F_{2}^{s}}{5})$ for 
right and left
circularly polarized sound waves ($m_{J} =
\pm1$).   The Tsuneto function $\lambda(\omega,T)$  can be calculated 
from the energy  gap
\cite{Moo93}. The Fermi liquid parameterization of quasiparticle 
interactions is given in
terms of $F_{1,2}^{s}$.  Both $a_{2^{-}}$ and the Land$\acute{e}$
$g$-factor,
$g_{2^{-}}$, are predicted to depend on the strength of quasiparticle 
and $f$-wave pairing
interactions and consequently on temperature and pressure.  In this 
letter we present our
measurements of the Faraday rotation angle from which we determine the
$g$-factor.  From our results, along with accurate measurements of $a_{2^{-}}$
\cite{Dav06}, we are able to determine the $f$-wave pairing 
interaction strength.

\begin{figure}[b]
\centerline{\includegraphics[width=3.4in]{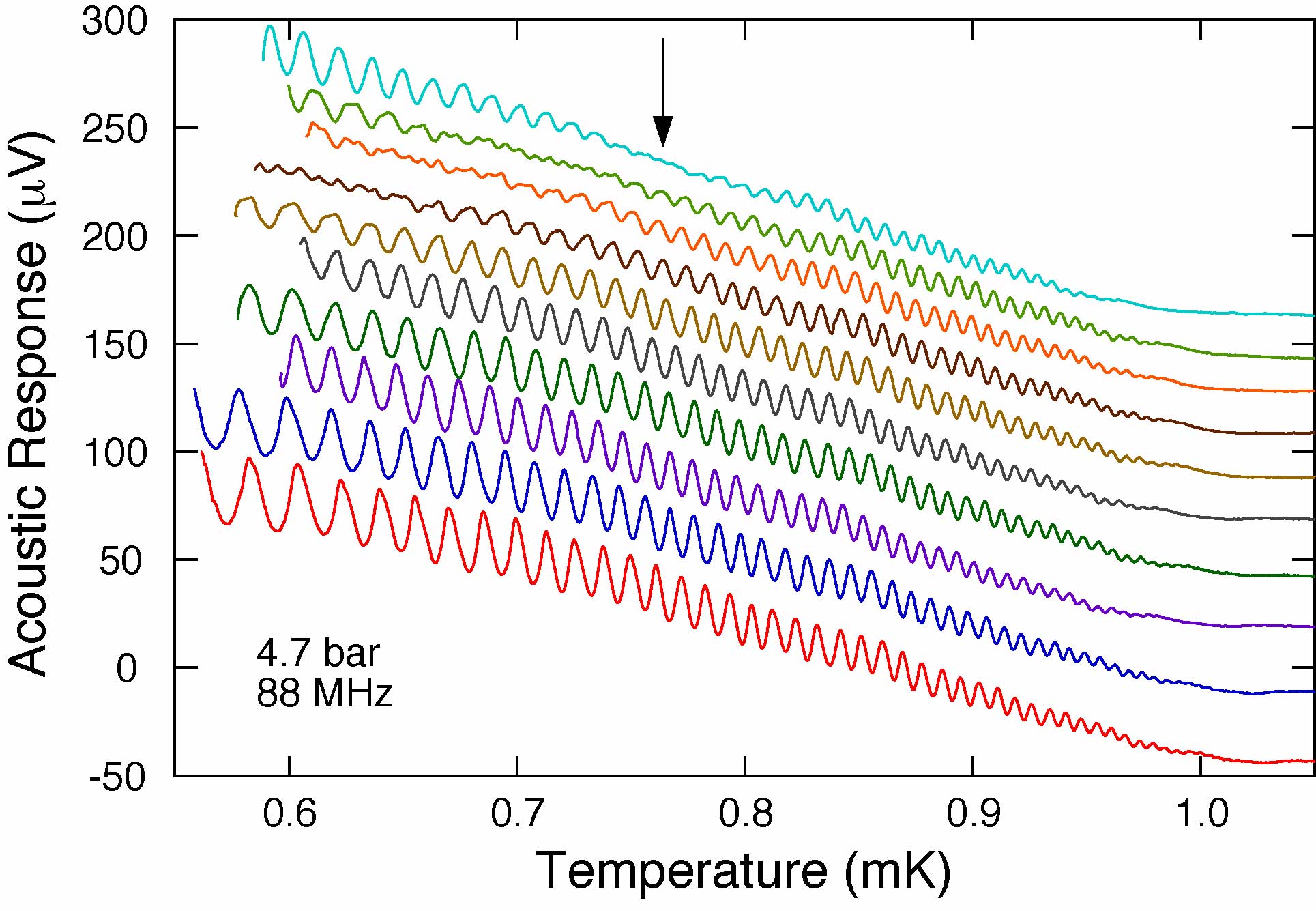}}
\caption {\label{fig2}(color online).  Acoustic cavity response as a 
function of
temperature for 4.7 bar and 88 MHz.  Each trace (offset for clarity) 
is for a different
magnetic field,  bottom to top: 0, 61, 91, 122, 152, 183, 244, 274, 
305 and 366 G.  As the
magnetic field is increased the linear polarization of transverse 
sound is rotated,
decreasing the amplitude of the  interference oscillations.  At 366 G 
it is clear that
there is a minimum (vertical arrow) in the acoustic response at 765
$\mu$K, which corresponds to a ${\pi/2}$ rotation of the polarization 
of the reflected
wave with respect to the  detection and generation direction.}
\end{figure}

We cool the liquid $^3$He by adiabatic nuclear demagnetization to 
temperatures $\approx$
500 $\mu$K. These methods,  including thermometry, are described 
elsewhere \cite{Ham89}.
A cavity for liquid
$^3$He is formed using an \emph{AC}-cut quartz transducer as one wall 
with an optically
polished quartz reflector as the other.  The spacing is defined by 
two wires with diameter
$\approx25~\mu$m.  Spring loading this cavity ensures that the cavity 
walls are parallel
and that the spacing is uniform over  the entire area of the cavity 
even as the experiment
is cooled to low temperatures.  The spacing was measured  at room 
temperature using
fluorescence techniques on a Zeiss Meta 500 confocal laser microscope 
at many places over
its area.  At 18 mK, $D = 31.6\pm 0.1~\mu$m was determined from the 
known dependence of
the  longitudinal sound velocity on pressure \cite{Hal90}.  We use overtone 
frequencies,
odd harmonics between 13 and 27, in the range 76 MHz to 159 MHz.

As the phase velocity of transverse sound changes so does the number 
of half-wavelengths
in the cavity, altering the acoustic impedance at the surface of the 
piezoelectric
transducer and producing a shift in the resonance spectrum which we 
detect with a
resolution of
$\approx 2\times10^{-6}$ using an
\emph{RF}-bridge, \emph{FM}-modulation, and lock-in  detection 
\cite{Ham89,Dav06}. In our
experiment we hold $\omega$ constant and sweep either the temperature 
or pressure to vary
the ISQ-mode frequency, $\Omega_{2^{-}}$ (see Fig.~\ref{fig1}).  We 
detect the changing
velocity as an oscillatory acoustic response, displayed in 
Fig.~\ref{fig2}. Precise
changes of the velocity can be measured and by
comparison with Eq.~\ref{dispersion} near the ISQ-mode \cite{Dav07}, we have
determined their absolute values.  We find that the  phase velocity 
approaches 500 m/s
close to the mode, so that it in this limit the first term in 
Eq.~\ref{dispersion},
$\Lambda_{0}$, is negligible.

It is expected \cite{Moo93} that off-resonant coupling of transverse 
sound to the ISQ-mode
holds only in the shaded region of  Fig.~\ref{fig1}.  Otherwise 
propagation is strongly
attenuated, either by pair breaking, $\omega > 2 \Delta$, or  if $\omega <
\Omega_{2^{-}}$.  In the latter case there are no real-valued 
solutions for $q$ in
Eq.~\ref{dispersion}.  In our previous work \cite{Dav06}, extended 
here to 27 bar, we used
the acoustic  signature for $\omega = \Omega_{2^{-}}(T,P)$ to obtain 
$\Omega_{2^{-}}(0,P)
= \sqrt{12/5}(1.0018 + 0.00144  P)\Delta^{+}(0,P) \pm 0.3\%$. These 
results are expressed
in terms of the weak-coupling-plus  gap \cite{Rai76,Hal90}, 
$\Delta^{+}(T,P)$, fixed to the
Greywall temperature scale \cite{Gre86}.  Additionally, we find that 
some harmonics of
\emph{AC}-cut transducers can generate {\it longitudinal} acoustic 
cavity resonances
giving oscillatory acoustic response immediately below the shaded region in
Fig.~\ref{fig1}.  We reported  this earlier \cite{Dav06} although, at 
the time, we were
unaware of its origin.

The acoustic Faraday effect is schematically represented in 
Fig.~\ref{fig1}. Application
of a magnetic field splits the ISQ-mode into  five components.  One of these,
$m_{J}=1$, couples to right circularly polarized sound and its 
frequency is increased by
field.  A second, $m_{J}= -1$, couples to left circular sound and is 
decreased by field.
Consequently, these circularly polarized waves have different phase 
velocities.  They
interfere and thus rotate the plane  of linear polarization 
proportional to path length
and to the applied magnetic field,  directly analogous to the 
magneto-optic effect
discovered by Michael Faraday.

\begin{figure}[t]
\centerline{\includegraphics[width=3.4in]{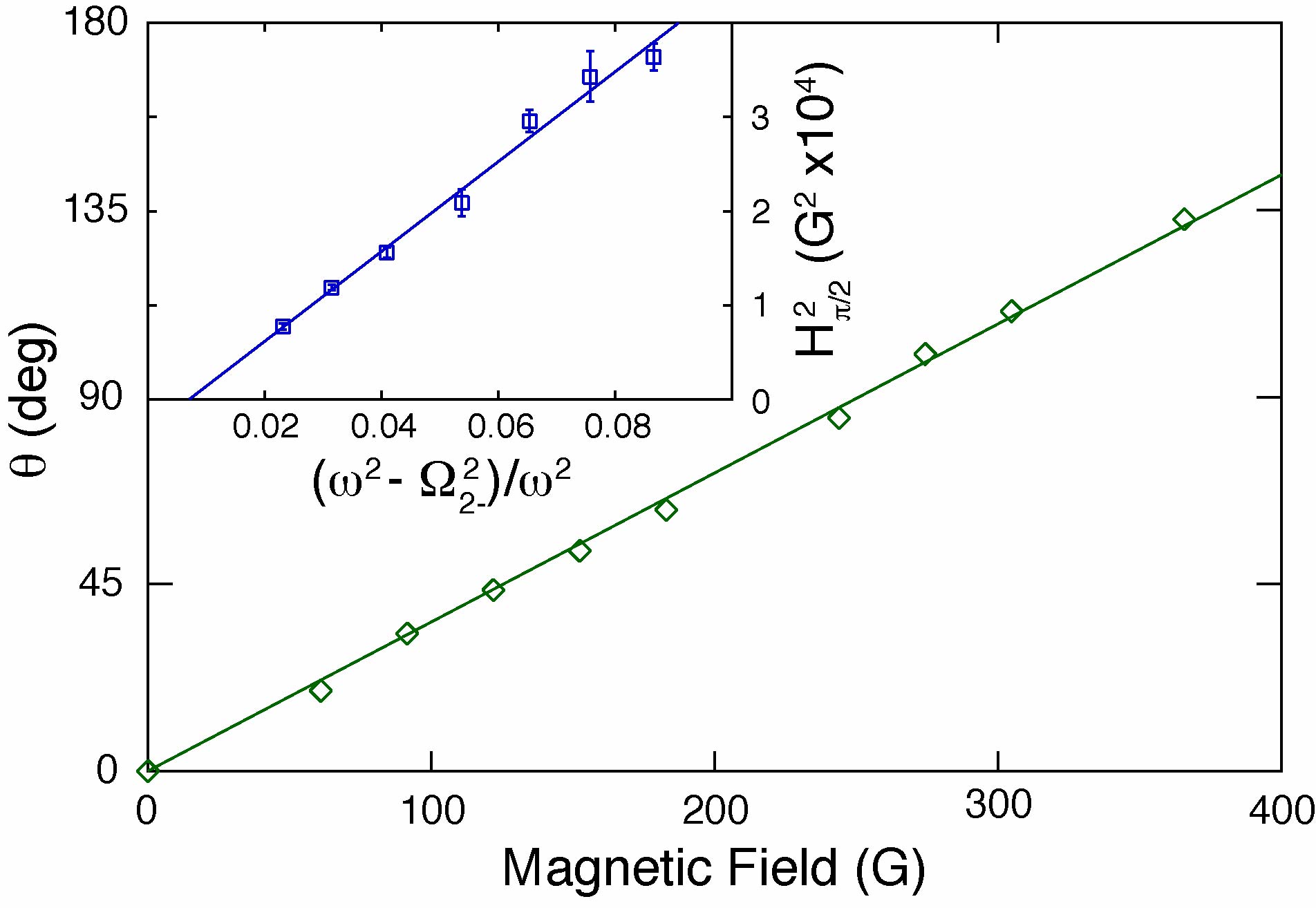}}
\caption {\label{fig3}(color online).  The Faraday rotation angle as 
a function of applied
field $H$ for representative data at $P= 4.7$ bar, $T= 626~\mu$K and 
frequency 88 MHz.
(Inset) The square of the field for a $\pi/2$ Faraday rotation angle, 
$H_{\pi/2}^2$,  is
shown as a function of proximity to the ISQ-mode, 
$\Omega_{2^{-}}(P)$, extrapolated to
$T=0$; with a frequency of 135.1 MHz and centered around 17.3 bar.}
\end{figure}

To accurately measure the AFE rotation angle, we slowly sweep either 
the temperature or
pressure at constant frequency in various magnetic fields applied 
along the direction of
propagation.  For temperature sweeps this corresponds  to the 
horizontal line in
Fig.~\ref{fig1}, typically in the range of $\approx$ 500 to 900 
$\mu$K.   Acoustic
response data for temperature sweeps in different magnetic fields are shown in
Fig.~\ref{fig2},  which we can represent in the form,
\begin{equation}\label{acousticimpedance} A = A_{0} + A_{1} \cos\theta
\sin(2D\omega/c + \phi),
\end{equation} where $\theta = \frac{\pi}{2}(H/H_{\pi/2})$ is the 
Faraday rotation  angle,
$H_{\pi/2}$ is the field that rotates the polarization by $\pi/2$ and 
$\phi$ is a phase.
The $\cos\theta$  factor is the projection of the rotated 
polarization with respect to a
fixed direction of polarization for generation and  detection of 
transverse sound,
characteristic of the transducer.  $A_{0}$ is a smoothly varying 
background signal in the
absence of  acoustic cavity  interference;  $A_{1}$ is the maximum 
signal modulation from
acoustic interference in the cavity. The rapid oscillatory  behavior 
in Fig.~\ref{fig2}
comes from the temperature dependence of the phase velocity, 
modulated by the field
dependence of  the Faraday rotation angle.  Typical results are shown 
in Fig.~\ref{fig3}
for $P= 4.7$ bar, $T= 626~\mu$K and a frequency of 88 MHz.

The amplitude of the oscillations in the acoustic response in 
Fig.~\ref{fig2}  decreases
as the magnetic field is increased at fixed temperature, passing 
through a minimum
indicated, for example, by  an arrow in Fig.~\ref{fig2} at 366 G and 
765 $\mu$K. This
corresponds to a
$\pi/2$ rotation of the linear polarization as the wave traverses a 
round-trip path in
the cavity.  From the data in Fig.~\ref{fig2} we find $\theta$ at 
constant temperature and
plot its dependence on magnetic  field in Fig.~\ref{fig3}. We find 
that the rotation angle
is proportional to magnetic field for all pressures.

It is convenient to work at the lowest temperatures to minimize 
temperature dependences,
so we extrapolate our measurements of $H_{\pi/2}(T)$ to
$T=0$ using a phenomenological expression that fits our data well,
$H_{\pi/2}(T) = H_{\pi/2}(0) + Be^{-2\Delta(0)/k_{B}T}$, where
$B$ is a fit parameter. An example of these results, inset to  Fig.~\ref{fig3},
shows that for acoustic frequencies approaching the ISQ-mode,
$H_{\pi/2}(0)$ becomes smaller, increasing the Faraday rotation angle in a
fixed field.

Quantitative comparison with theory \cite{Sau99} requires that we 
calculate the
$g$-factor in terms of $H_{\pi/2}$, using the condition for $\pi/2$ rotation,
$q_{+}- q_{-} = \frac{\pi}{2D}$ where $q_{\pm}$ corresponds to $m_{J} 
=\pm1$  in Eq.1.
We show that sufficiently close to the ISQ-mode,
\begin{equation}\label{g} g_{2^{-}} =
\sqrt{\frac{\omega^2-\Omega_{2^{-}}^{2}}
{\omega^2}}\frac{v_F\Lambda_{2^{-}}^{1/2}}{\gamma_{eff}H_{\pi/2}8D}.
\end{equation}  Determination of
$g_{2^{-}}$ depends on  precise knowledge of the ISQ-mode frequency,
$\Omega_{2^{-}}$ as defined above, which we have measured 
independently in zero field.
This formulation for magneto-aoustics  follows from the dispersion relation,
Eq.~\ref{dispersion}, which can be verified  experimentally.   We 
have performed seven
measurements of $H_{\pi/2}(0)$ at a single acoustic frequency of 
135.1 MHz within a small
pressure range from $P=16.4$ to  18.2 bar, that tunes the ISQ-mode 
frequency from 129.1 to
133.5 MHz (over which the variation in
$g_{2^{-}}$ is small).  Our results in the inset of Fig.~\ref{fig3} 
are a  validation of
the predicted frequency behavior in Eq.~\ref{g}. An unconstrained 
linear fit to the data
has a small offset that  corresponds to a shift in ISQ-mode frequency 
of 0.3\% which is
within experimental error.

\begin{figure}[b]
   \centerline{\includegraphics[width=3.4in]{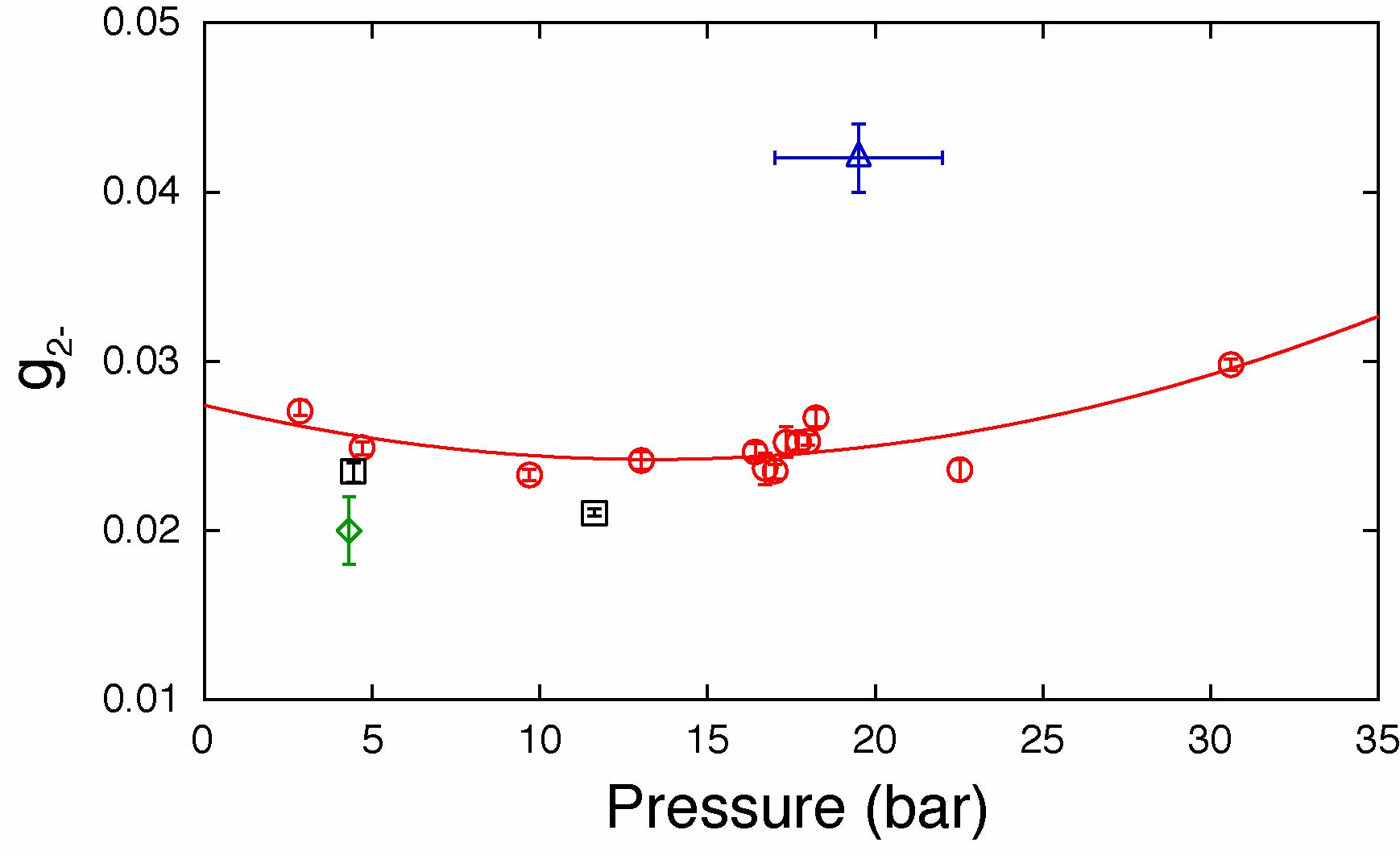}}
\caption {\label{fig4}(color online). The Land$\acute{e}$ $g$-factor,
$g_{2^{-}}$, as a function of pressure at $T=0$.   The red circles 
are from Faraday
rotation measurements  as a function of temperature, extrapolated to 
zero temperature.
The black squares are measurements performed via pressure sweeps at our lowest
temperatures.  The blue triangle is from Movshovich
\emph{et al.} \cite{Mov88} and the green diamond is from Lee \emph{et 
al.} \cite{Lee99}.
The curve is given by $g_{2^-}=0.0274-4.8\times10^{-4}P+1.8\times10^{-5}P^2$.}
\end{figure}

Eq.~\ref{g} is accurate when the difference between the acoustic 
frequency and the
ISQ-mode frequency is small, which is valid for some of our data. 
Outside of this
limit the quasiparticle term, $\Lambda_0$, as well as the dispersion term,
$\frac{2}{5}q^2v_{F}^2$, become non-negligble and must be taken into account.
Accordingly, we have calculated $g_{2^{-}}$ from $H_{\pi/2}(0)$ using 
the full dispersion
relation in Eq.~\ref{dispersion}, which is solved numerically.   This 
is performed for all
of our data and the resulting $g_{2^{-}}$ values are presented in 
Fig.~\ref{fig4}.

A second method for data acquisition is to sweep the pressure at our 
lowest possible
temperature, $\approx$ 500 $\mu
$K, for various applied magnetic fields. In this case the pressure 
dependence of the
transverse velocity is responsible for  oscillatory acoustic response 
similar to that
shown for temperature sweeps, as in Fig.~\ref{fig2}.   From these 
measurements we
obtain
$H_{\pi/2}(0)$ and, as described above, determine the values for 
$g_{2^{-}}$ shown as
black squares in Fig.~\ref{fig4}. There is good  agreement between 
the two methods.

\begin{figure}[t]
   \centerline{\includegraphics[width=3.4in]{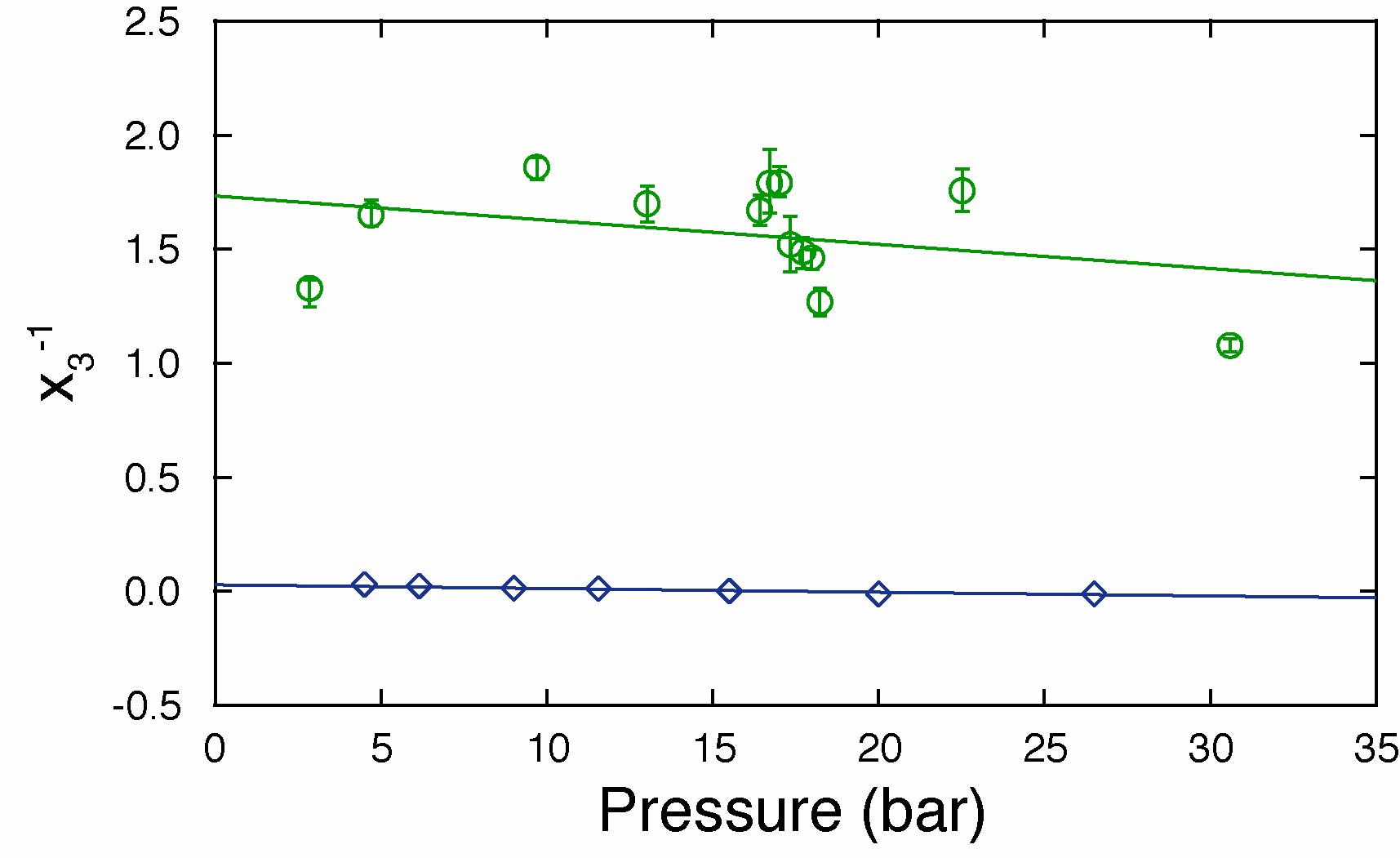}}
\caption {\label{fig5}(color online). The green circles are the 
$f$-wave interaction
strength $x_3^{-1}$ as function of pressure obtained from measurments 
of the ISQ-mode
$g$-factor and frequency measurements and the blue diamonds are from 
ISQ-mode frequency
measurements alone \cite{Dav06}.  The solid lines are guides to the eye. The
significant difference betwen these two data sets is likely associated with
strong coupling effects on the ISQ-mode frequencies, discussed in the text.}
\end{figure}

Our results for $g_{2^{-}}$ cover the full pressure range of  superfluid
$^3$He-B.  The theoretical value for $g_{2^{-}}(T=0)$ without the 
effects of $F^{s}_{2}$
or $f$-wave interactions is 0.0372 \cite{Sch81,Sau82}.  A previous 
result, shown as a
green diamond in Fig.4,  by Lee {\it et al.} \cite{Lee99} from the AFE at
4.4 bar, agrees qualitatively with our work, where the difference can 
be ascribed to less
accuracy in their determination of
$\Omega_{2^{-}}$.  Movshovich {\it et al.} \cite{Mov88} used longitudinal
acoustics at much higher magnetic fields to directly determine the 
ISQ-mode splitting.
Their considerably higher result, shown as a blue triangle, is likely
due to non-linear magnetic field dependence of the mode frequencies 
at large magnetic
fields.

The deviation of
$g_{2^{-}}$ from its from weak coupling value is an indication of the 
important role of
$^3$He quasiparticle and pairing interactions.  According to the 
theory of Sauls and
Serene
\cite{Sau82} the two relevant parameters are the Landau parameter
$F_2^s$, and the  strength of $f$-wave pairing interactions,
$x_3^{-1} = 1/(v_1^{-1} -v_3^{-1})$. Here $v_1$ and $v_3$ are the 
pairing potentials due
to $p$-wave and $f$-wave interactions respectively.   The theory 
allows calculation of
$x_3^{-1}$ from $g_{2^{-}}$ and the zero-field ISQ-mode frequencies, 
with $F_2^s$ and the
energy gap as inputs.  We use values for $F_2^s$, based primarily on 
measurements of the
difference between zero- and first-sound velocities in the normal 
fluid, and $\Delta^+$ as
tabulated in Ref.~\cite{Hal90}.  The theoretical expression for 
$g_{2^{-}}$ is complex
and is not reproduced here.  The green circles in Fig.~\ref{fig5} 
show the result of this
calculation, with all predicted dependences on these parameters 
including the explicit
dependences of $\Lambda_{2^{-}}$ and $\Lambda_0$ on
$F_2^s$.

We find that $x_3^{-1}$ is positive, which means that $f$-wave 
interactions  are
repulsive, with little pressure dependence.  These values are in 
disagreement with
our results  from analysis of the ISQ-mode frequencies \cite{Dav06},
interpreted within the framework of Ref.~\cite{Sau81}, shown as blue 
diamonds in
Fig.~\ref{fig5}.  We believe  that the discrepancy between these two 
calculations of
$x_3^{-1}$ likely originates in non-trivial strong coupling 
corrections to the ISQ-mode
frequencies.  For example, at 5 bar our measurements reveal that 
Fermi liquid and $f$-wave
corrections to the mode frequencies are
$\lesssim 0.9\%$, whereas strong coupling corrections to $\Delta$ are 
$2.5\%$.  This
situation persists at all pressures, suggesting that calculations of the mode
frequencies beyond the weak-coupling-plus model \cite{Rai76} are 
required.  On the other
hand the
$g-$factor is more strongly modified by interaction effects as 
compared with the
mode-frequency and so $x_3^{-1}$ derived from the $g-$factor should 
be more robust.
Variation of the absolute temperature scale by 1 \% and $F_2^s$ by 
0.5 have negligible
effect on the calculation of the $g$-factor and change $x_3^{-1}$ 
within the scatter
of the data.

One of the predictions of Sauls and Serene \cite{Sau81} is that for
some combination of either negative values of
$x_3^{-1}$ or a negative higher order Landau interaction term,
$F_4^s$, there is a new order parameter collective mode close to the 
gap edge that
corresponds to total  angular momentum $J=4$.  Longitudinal sound 
attenuation measurements
\cite{Lin87} near $\omega = 2\Delta$ were  initially interpreted in 
this way although this
was later revised as providing evidence for a $J=1$ collective mode. 
Acoustic Fourier
transform spectrosopy \cite{Mas00} indicated  an unexplained 
additional contribution to
attenuation near the gap and an explanation in terms of the $J=4$ 
mode  was also
considered. It should be possible to use  high resolution acoustic 
cavity methods, as in
the present work, to look for these modes although our determination 
of the repulsive
interaction in the $f$-wave channel is not encouraging.

In summary, we have investigated the magneto-acoustic Faraday effect 
in superfluid
$^{3}$He using high resolution transverse sound techniques.  Our 
spectroscopic data for
the ISQ-mode  frequency and its $g$-factor are the most complete 
characterization
of an order parameter collective mode in superfluid
$^{3}$He.  We find from our results, combined with the theory, that 
$f$-wave pairing
interactions are repulsive.  But strong-coupling corrections to the 
ISQ-mode frequencies
may be required to understand the full effect of
$f$-wave interactions on the collective modes.

We acknowledge support from the National Science Foundation,
DMR-0703656 and thank J.A. Sauls, T.M. Lippman, W.J. Gannon, and B. 
Reddy for useful
discussions and Y. Lee and G. Gervais for early contributions.

\end{document}